\documentclass[aps,longbibliography,preprint,%
 superscriptaddress, 
 amsmath,amssymb, nofootinbib,       
 floatfix,
]{revtex4-2}

\usepackage{graphicx}
\usepackage{dcolumn}
\usepackage{bm}
\usepackage[dvipsnames]{xcolor}
\usepackage{hyperref,url}
\usepackage{cleveref}
\usepackage{orcidlink}
\usepackage{shuffle}
\usepackage{graphicx}
\usepackage{physics}
\usepackage{array}
\newcolumntype{P}[1]{>{\centering\arraybackslash}p{#1}} 
\usepackage{tikz}

\newcommand{\nn}{\nonumber \\}

\newcommand{\pol}{\ensuremath{\varepsilon}}
\newcommand{\levi}{\ensuremath{\epsilon}}

\begin{document}

\title{Entire four-graviton EFT\\
from the duality between color and kinematics}

\author{John Joseph M. Carrasco \orcidlink{0000-0002-4499-8488}}
\affiliation{Amplitudes and Insights Group, Department of Physics and Astronomy, Northwestern University, Evanston, IL 60208, USA}
\affiliation{Center for Interdisciplinary Exploration and Research in Astrophysics (CIERA), Northwestern University, 1800 Sherman Ave, Evanston, IL 60201, USA}
\author{Sai Sasank Chava \orcidlink{0009-0009-0228-9737}}
\affiliation{Amplitudes and Insights Group, Department of Physics and Astronomy, Northwestern University, Evanston, IL 60208, USA}
\author{Alex Edison \orcidlink{0000-0002-5430-9500}}
\affiliation{Amplitudes and Insights Group, Department of Physics and Astronomy, Northwestern University, Evanston, IL 60208, USA}
\affiliation{Center for Interdisciplinary Exploration and Research in Astrophysics (CIERA), Northwestern University, 1800 Sherman Ave, Evanston, IL 60201, USA}
\author{Eliseu Kloster \orcidlink{0009-0002-1796-1628}}
\affiliation{Department of Astronomy and Astrophysics, University of Chicago, Chicago, IL 60637, USA}
\affiliation{Amplitudes and Insights Group, Department of Physics and Astronomy, Northwestern University, Evanston, IL 60208, USA}
\affiliation{Center for Interdisciplinary Exploration and Research in Astrophysics (CIERA), Northwestern University, 1800 Sherman Ave, Evanston, IL 60201, USA}
\author{Suna Zekioğlu \orcidlink{0000-0002-5019-7310}}
\affiliation{Amplitudes and Insights Group, Department of Physics and Astronomy, Northwestern University, Evanston, IL 60208, USA}

\begin{abstract}
The Bern-Carrasco-Johansson (BCJ) double-copy construction reveals a fundamental structural connection between gauge and gravity theories. At its core, the BCJ double copy is directly due to a duality between the algebraic relations of a color root and those of a kinematic root. We generalize this principle beyond the conventional Lie algebra structure of tree-level Yang-Mills theory. By demanding color-kinematics duality for the complete basis of four-point color structures --- including those involving the symmetric $d^{abc}$ constants --- we define the universal double copy. We systematically classify the bases of all such parity-even generalized gauge-theory numerators and, independently, the space of all parity-even four-graviton higher-derivative operators. We demonstrate that our universal double-copy construction precisely spans the entire tower of parity-even four-graviton amplitudes in any dimension, except for the Lovelock $R^3$ contribution in $D >6$ which we can express in terms of a particularly simple universal triple-copy involving gauge theories coupled to scalars. Explicit machine-readable expressions for the complete basis of gauge-theory numerators and fundamental gravitational building blocks are provided in the ancillary files.  This establishes that all possible four-point gravitational interactions can be factorized into products of gauge-theory building blocks governed by this universal notion of color-kinematics duality.
\end{abstract}

\maketitle
\newpage

\tableofcontents

\section{Introduction}
\label{sec:introduction}

The principle of the correspondence between color and kinematics, and the associated double-copy construction, posits a deep and predictive relationship between the gauge-invariant content of gauge and gravity theories~\cite{Bern:2008qj,Bern:2010ue}.  While the factorization of gravity into gauge theory was first identified in the Kawai-Lewellen-Tye (KLT) relations~\cite{Kawai:1985xq} and their field-theory limits~\cite{Berends:1988zp,Bern:1998sv}, the modern duality identifies a strictly local graph-based organization.   At its core lies the realization: at tree-level gauge-theory amplitudes admit representations in which their kinematic numerators can be arranged graph by graph to satisfy the same algebraic relations as the corresponding color factors. This correspondence extends beyond scattering amplitudes, linking the double-copy web of theories (including gravity) at the level of their local interaction quantum operators --- the vertices appearing in the expansion of the gauge-invariant effective action~\cite{Bern:1999ji,Carrasco:2025ymt}.

This graph-based organization has extended the reach of perturbative prediction.
It has enabled the construction of multiloop scattering amplitudes in
supergravity at orders previously thought impossible, offering critical insights
into the ultraviolet properties of quantum
gravity~\cite{Bern:2007hh,Bern:2009kd, Bern:2012cd, Bern:2012uf, Bern:2013uka,
Bern:2014sna, Bern:2018jmv}. In parallel, the formalism has become an
increasingly pivotal tool for calculating classical observables relevant to
gravitational-wave astronomy. Extracting the classical limit of scattering
amplitudes in quantum field theory (QFT)~\cite{Cheung:2018wkq, Kosower:2018adc} offers a highly efficient
route to state-of-the-art post-Minkowskian interaction potentials~\cite{
Bern:2019nnu, Bern:2019crd, Bern:2020buy, Bern:2021dqo, Bern:2022kto}. These QFT
amplitude-based advances are synergistic with worldline effective field theory
insights~\cite{Goldberger:2004jt, Mogull:2020sak, Jakobsen:2021smu,
Jakobsen:2021lvp,Edison:2022cdu,Edison:2023qvg,Edison:2024owb}, where compatible double-copy structures have notably been
identified in the description of classical radiation~\cite{Shi:2021qsb,Almeida:2020mrg}.

Importantly, the scope of color-kinematics duality extends beyond ordinary
perturbation theory. In tree-level string theory, all-order $\alpha'$-resummed
amplitudes admit color-dual representations, allowing perturbative string
amplitudes to be understood as double copies between point-particle gauge
theories and nonlocal effective field theories that encode the effects of an
infinite tower of massive
modes~\cite{Mafra:2011nv,Mafra:2011nw,Broedel:2013tta,Carrasco:2016ldy,Carrasco:2016ygv,Mafra:2016mcc,Azevedo:2018dgo,Edison:2021ebi,Edison:2022smn}.
More recently, color-kinematics duality has also been shown to organize
genuinely nonperturbative phenomena in $\hbar$: the thermal structure of
Yang-Mills pair production admits a color-dual description whose double copy
reproduces the thermality of Hawking radiation~\cite{Carrasco:2025bgu}.

In all these settings, replacing color weights by kinematic weights promotes gauge invariance in the single-copy theory to diffeomorphism invariance in the gravitational theory, ensuring the correct physical properties of the resulting amplitudes, including statistics and factorization~\cite{Chiodaroli:2017ngp,Bern:2019prr}.

Since its inception, a central question has been to determine the ultimate scope of this principle. The original formulation focused on the Lie algebra structure of antisymmetric adjoint color factors ($f^{abc}$), which, while remarkably successful in relating a wide web of theories, does not encompass the full landscape of possible effective field theory (EFT) operators~\cite{Mafra:2011nv,Mafra:2011nw,Broedel:2012rc, Boels:2016xhc,Elvang:2020kuj,Carrasco:2022jxn,Carrasco:2023qgz}. This points toward a necessary generalization: the duality principle should apply not just to the antisymmetric-adjoint sector, but to the complete algebra of color structures available to a gauge theory.

Our recent work has pursued this generalization systematically. The principle of
spanning complex amplitudes from a small basis of kinematic building blocks was
first established in our work spanning higher derivative corrections to
bi-colored scalar theories at four~\cite{Carrasco:2019yyn} and
five-points~\cite{Carrasco:2021ptp} allowing for a mixture of arbitrary color
and higher derivative scalar-kinematics to be antisymmetric-adjoint color-dual.
This spans, e.g.,  the Chan-Paton ordering of the $Z$-theory amplitudes~\cite{Broedel:2013tta, Carrasco:2016ldy,Carrasco:2016ygv, Mafra:2016mcc} which lift pointlike quantum field theories to string theories via field-theory double-copy.   However, it became clear that the set of purely antisymmetric-adjoint building blocks was insufficient to capture all physical effective operators. This led us to demonstrate the necessity of a symmetric double copy, based on a kinematic algebra dual to symmetric color structures ($d^{abc}$), to construct operators inaccessible to the traditional framework~\cite{Carrasco:2022jxn,Carrasco:2023qgz}. These successes in identifying and utilizing a broader set of building blocks and algebraic rules culminated in the development of a comprehensive, modular framework for systematically constructing all such gauge-theory amplitudes~\cite{Carrasco:2025pmi}.

The present paper brings this program to a definitive conclusion at four points.
We present a  systematic and complete classification of all
parity-even\footnote{We consider parity even specifically because it is
dimension agnostic ---  the same techniques can be brought to bear to consider
spacetime parity odd in any particular dimension.} , four-graviton amplitudes at
any mass dimension in general spacetime, derived from the first principles of
symmetry and locality. We then demonstrate that this entire space is precisely
spanned by gauge theory predictions via the universal $N$ copy: a
construction that synthesizes the kinematic algebras dual to all possible
four-point adjoint color structures and generalizes beyond simply two copies. We
demonstrate that while the $N = 2$ (double) copy spans the vast majority of the
landscape, including all higher-derivative four-point graviton amplitudes in $D
\leq 6$, the $R^3$ Lovelock term relevant to four-points requires a generalized
$N=3$ (triple)-copy structure in $D > 6$.  We expect that the full tower of
Lovelock terms requires the full generalization to arbitrary $N$-copies each
relevant starting at multiplicity $N+1$. 

\section{The Universal $N$-Copy at Four Points}
\label{sec:universal_dc}

To realize the full scope of the color-kinematics duality, we begin by identifying the complete set of algebraic relations that color-factors can satisfy at four-points, and then demand corresponding relations for the notion of kinematic duals to such color structures.   The traditional framework emphasizes antisymmetric-adjoint color structures built out of $f^{abc}$ structure constants and thus obey Jacobi and antisymmetry relations.  While remarkably powerful, this algebraic structure does not exhaust all ways of consistently dressing cubic graphs at four points.

Generally one may consider color-factors with different symmetry properties at each vertex, including structures that are symmetric rather than antisymmetric, and which satisfy no Jacobi-like relations.  These define distinct algebraic sectors for dressing cubic graphs.  Additionally one may dress contact graphs.  The principle of color-kinematics duality suggests that for each of these algebraic sectors  there should exist a corresponding kinematic realization obeying the same relations. The universal double copy is then defined as the synthesis of constructions built from each of these sectors. We now detail the algebraic properties, double-copy rules, and KLT-like structures for each sector in turn.

Our use of the symbols $f^{abc}$ and $d^{abc}$ in what follows is purely notational: they label distinct \emph{vertex symmetry types} and the corresponding algebraic identities satisfied by four-point color factors (antisymmetric, mixed-symmetry, and symmetric).  The universal double-copy constructions are defined by these algebraic relations alone. 

Along these lines,  we will introduce the notion of color-ordered amplitudes. Throughout this work by ordered (or partial) amplitudes we mean the gauge-invariant kinematic coefficients associated with a chosen independent color basis.  When convenient --- particularly to contrast with familiar KLT-type expressions ---  we will specialize to the trace basis of $SU(N_c)$ as an explicit realization, but the underlying constructions are defined independently of any particular gauge-group identification.

We note that our formulation of KLT-like structures in the following subsections for kinematics dual to color structures involving $d^{abc}$ and $d^{a_1 a_2 a_3 a_4}$ is distinct from  that of e.g.~refs.~ \cite{Chi:2021mio,Chen:2023dcx}.  The KLT-like kernels we consider do not necessarily conform to the traditional ($ff$) KLT kernel, which has a rank of $(n-3)!$.  Instead the rank of such a matrix depends on the color-structure under consideration. We frame our discussion of universal double copies for arbitrary color-structures both at the level of color-dual numerators and the analogs of $f^{abc}$ color-ordered amplitudes. The prescription of double copy in terms of the numerators makes the pole structure directly manifest. 

As such, when local dual representations are identified that satisfy
color-kinematics duality they do not suffer from the issues of spurious poles discussed in Refs.~\cite{Chi:2021mio,Chen:2023dcx}, although they can incur additional operator content\footnote{Maintaining the ``purity'' of the original theory (not bootstrapping towards additional operators required for color-dual locality), but instead exploiting nonlocal color-dual numerators can of course introduce spurious poles.} than initially posed for the theory~\cite{Carrasco:2022lbm, Carrasco:2022sck, Brown:2023srz, Li:2023wdm, Carrasco:2023wib}.  Consistently specifying the color-dual theory carves a space in effective field theories.  We can obtain the universal double-copy prescription at the level of color-ordered amplitudes by inverting the KLT-like kernel that relates the amplitudes to the numerators. As such, the healthy pole behavior of universal double copy continues to hold at the level of amplitudes as well.

Throughout this paper we will label cubic (trivalent) graphs clockwise starting from the bottom-leftmost label,
\begin{equation}
 n(i j;k l) = n \left( 
 \begin{tikzpicture}[baseline=(c), scale=0.8, line width=0.9pt]
  \coordinate (vL) at (-1.0, 0.0);
  \coordinate (vR) at ( 1.0, 0.0);

  \coordinate (p1) at (-2.4,-1.2); 
  \coordinate (p2) at (-2.4, 1.2); 
  \coordinate (p3) at ( 2.4, 1.2); 
  \coordinate (p4) at ( 2.4,-1.2);

  \draw (vL) -- (vR);

  \draw (vL) -- (p1) node[pos=1, left=2pt] {$i$};
  \draw (vL) -- (p2) node[pos=1, left=2pt] {$j$};

  \draw (vR) -- (p3) node[pos=1, right=2pt] {$k$};
  \draw (vR) -- (p4) node[pos=1, right=2pt] {$l$};

  \coordinate (c) at (0,0);
\end{tikzpicture} \right)\,.
\end{equation}
In addition, we will often use the shorthands
\begin{equation}
n_s = n(12;34)\,, \quad n_t = n(14;32)\,, \quad n_u = n(42;31)\,.
\end{equation}

\subsection{The antisymmetric-adjoint ($ff$) sector}
\label{sec:antisymFF}
The traditional BCJ double copy corresponds to the $ff$ sector, whose color factors,
\begin{equation}
c^{ff}_s = f^{a_1 a_2 e_s}f^{e_s a_3 a_4}~, ~~~ c^{ff}_t = f^{a_1 a_4 e_t}f^{e_t
a_3 a_2}~, ~~~c^{ff}_u = f^{a_4 a_2 e_u}f^{e_u a_3 a_1}\,,
\end{equation} 
satisfy the Jacobi identity, $c^{ff}_s - c^{ff}_t - c^{ff}_u = 0$. A gauge theory is color dual in this sector if its kinematic numerators obey the same identities.  We list the following minimum relations necessary for a $n^{ff}(i j;k l)$ graph to be considered color-dual,
\begin{align}
    \label{eq:ffColorDual_anti}   n^{ff}(i j;k l) &= - n^{ff}(j i;k l) & &\text{antisymmetry}\,, \\
    \label{eq:ffColorDual_iso}   n^{ff}(i j;k l) &= n^{ff}(k l;i j) &
                                                 &\text{left/right isomorphism}\,, \\
    \label{eq:ffColorDual_Jacobi} n^{ff}_s -n^{ff}_t -n^{ff}_u  &= 0 & &\text{Jacobi}\,.
\end{align}
Antisymmetry and left/right isomorphism need only be satisfied functionally
under relabeling when little-group weight is shared by exchanged particles.  

There are additional functional isomorphisms which relate $n^{ff}_s$,
$n^{ff}_t$, and $n^{ff}_u$ when the little-group weight admits it.
For the all-vector cases considered in this and the subsequent section the three dressings are indeed automorphic, and satisfy:
\begin{align}
 n^{ff}_t &= \left. n^{ff}_s \right |_{2\leftrightarrow 4} = n^{ff}(14;32) \\ 
 n^{ff}_u &=\left. n^{ff}_s \right |_{1\leftrightarrow 4}=  n^{ff}(42;31)\,.
 \end{align}
However, one could also have a ($ff$) color-dual representation of the scattering
between two gluons and two scalars.  If the scalars are legs labeled $2$ and
$3$, then $n_t \neq n_s |_{k_2 \leftrightarrow k_4}$.   The replacement rule is
trying to functionally exchange a leg with little-group weight 0 with one that
has little-group weight 2 which is meaningless.  Rather the graph associated
with the $t$ channel is understood to be distinct from the graph associated with
either the $s$ or $u$ channel, which are related under relabeling.   

The double-copy amplitude is then given by
\begin{align}\label{eq:ff-double-copy}
    \mathcal{M}^{ff}(1234)
    &= \sum_{i \in \{s,t,u\}} \frac{\tilde{n}^{ff}_i n^{ff}_i}{P_i}\,,
\end{align}
where $n^{ff}$ and $\tilde{n}^{ff}$ are two sets of color-dual numerators, and the propagators follow their labels, e.g.~$P_s =s$. This can be expressed via the well-known KLT relations~\cite{Kawai:1985xq,Bern:1998sv}  in terms of color-ordered amplitudes --- the gauge invariant kinematic coefficients of color-weights --- when expressed in a minimal basis,
\begin{align}\label{eq:ff-KLT}
    \mathcal{M}^{ff}(1234) = s_{13} A^{ff}(1234)\tilde{A}^{ff}(1324)\,.
\end{align}
Indeed one of the virtues of the correspondence between color and kinematics is that it allows a direct derivation of field-theory relations between ordered amplitudes, as well as momentum-kernel-type relations when double copying in terms of those amplitudes~\cite{Bern:2008qj}.  Pseudoinverting the matrix relating ordered
amplitudes and color-dual numerators yields color-dual numerators in terms of ordered amplitudes.  It is a pseudoinversion when there are nontrivial relations between ordered amplitudes yielding color-dual relations like the BCJ relations for $n^{ff}$.   Double copy of amplitude-expressed numerators yields a momentum kernel or KLT-like double-copy expression between ordered amplitudes\cite{Stieberger:2009hq,Bjerrum-Bohr:2009ulz,Bjerrum-Bohr:2010pnr,Mizera:2016jhj,Chi:2021mio,Chen:2023dcx} . 
This naturally allows the generalization of  KLT-like structure to algebraic
relations beyond adjoint-antisymmetric $c^{ff}$, as, e.g., was first
noted~\cite{Bargheer:2012gv, Huang:2012wr, Huang:2013kca} in the case of
three-dimensional Bagger-Lambert-Gustavsson theory~\cite{Bagger:2007jr}. 

\subsection{The mixed-symmetry ($df$) sector}

The first generalization involves color factors with mixed symmetry, such as $c^{fd}(1234) = f^{12e}d^{e34}$ and $c^{df}(1234) = d^{12e}f^{e34}$. In addition to the usual (anti)symmetric relations under exchanges of vertices, these factors obey the Jacobi-like algebraic relation~\cite{Bilal:2001hb} 
\begin{align}\label{eq:df-color-rel}
    c^{df}(1234)+c^{df}(4231)+c^{df}(1432)&=0.
\end{align}

So the minimum set of relations an $n^{df}$ color-dual to $c^{df}$ must satisfy are as follows,
\begin{align}
\label{eq:dfColorDual_anti}   n^{df}(i j;k l) &= +n^{df}(j i;k l) \,, \\
\label{eq:dfColorDual_sym}   n^{df}(i j;k l) &= -n^{df}(i j;l k) \,, \\
\label{eq:dfColorDual_Jacobi} n^{df}_s +n^{df}_t +n^{df}_u  &= 0 \,.
\end{align}
The numerators $n^{fd}$, which dress the color factors $c^{fd}$, must be related to $n^{df}$ by
\begin{align}
\label{eq:df-fd}   n^{df}(i j;k l) &= +n^{fd}(k l;i j) \,.
\end{align}
Any isomorphism relations between the numerators can only be satisfied functionally for same-type particles, as was the case for $ff$ numerators. Additional particle types will require potentially distinct graphs between the three channels and consequently distinct functional forms for isomorphic dressings. Again for the all vector dressings considered here and the subsequent section, isomorphic dressings are possible which satisfy:
\begin{align}
 n^{df}_s &= n^{df}(12;34) \,,\\
 n^{df}_t &=  \left. n^{df}_s \right |_{2\leftrightarrow 4}= n^{df}(14;32) \,, \\
 n^{df}_u &= \left. n^{df}_s \right |_{1\leftrightarrow 4}=  n^{df}(42;31) \, .
 \end{align}

A general amplitude in this sector takes the form:
\begin{align}\label{eq:df-amp}
    \mathcal{A}^{df}
    = \sum_{i \in \{s,t,u\}} \frac{c_i^{fd}n_i^{fd} + c_i^{df}n_i^{df}}{P_i}\,.
\end{align}
The double copy is then constructed by replacing the color factors with a second set of kinematic numerators, $\tilde{n}^{fd/df}$, that satisfy the same relations~\eqref{eq:df-color-rel}:
\begin{align}\label{eq:df-double-copy}
    \mathcal{M}^{df}
    = \sum_{i \in \{s,t,u\}} \frac{\tilde{n}_i^{fd}n_i^{fd} + \tilde{n}_i^{df}n_i^{df}}{P_i}\,.
\end{align}

It is worth specializing and considering the trace-basis relevant to $(S)U(N_c)$.  In this basis, expanding out the color-factor,
\begin{align}\label{eq:sym-constant}
    d^{abc}&=\text{Tr}\left(\{T^a,T^b\}T^c\right)\,.
\end{align}
and using standard color-trace identities  we can write the color-dressed gauge amplitude as 
\begin{align}\label{eq:df-amp-2}
    \mathcal{A}^{df }(1^a 2^b 3^c 4^d)
   &=\sum_{\sigma \in S_4/Z_4} \text{Tr} \left( T^{\sigma(1)} \dots T^{\sigma(4)} \right) A^{df}\left( \sigma \right),
\end{align}
where
\begin{align}\label{eq:df-color-ordered-amp}
 A^{df}(1234)=\frac{n^{df}(1234)+n^{fd}(1234)}{s}+\frac{n^{df}(2341)+n^{fd}(2341)}{t}.
\end{align}

The $df$ color-ordered amplitudes $A^{df}$ satisfy the modified color-reversed relation:
\begin{align}
    A^{df}(4321) = -A^{df}(1234).
\end{align}
As such, structures arising from $df$ dressings do not show up in $Z$-theory and open superstring theory.

The double-copy amplitude can be expanded over a basis of three independent partial amplitudes, such as $\{ A^{df}(1234), A^{df}(1243), A^{df}(1324) \}$. Following the same principle as in the adjoint sector, color-kinematics duality allows for the inversion of the relationship between these partial amplitudes and the color-dual numerators. This inversion yields a formula for the full amplitude in \cref{eq:df-double-copy} in terms of a color-dual momentum kernel, $\mathcal{S}^{df}$, acting on the partial amplitudes of the two single-copy theories:
\begin{widetext}
\begin{align}\label{df-KLT}
    \mathcal{M}^{df}(1234)
    &= A^{df} \cdot \mathcal{S}^{df}[s,t,u] \cdot \tilde{A}^{df\intercal} \nn
    &= \left(
\begin{array}{c c c}
 A^{df}(1234) &
 A^{df}(1243) &
 A^{df}(1324) \\
\end{array}
\right) \left(
\begin{array}{ccc}
 -\frac{st}{2u} & 0 & 0 \\
 0 & -\frac{su}{2t} & 0 \\
 0 & 0 & -\frac{tu}{2s} \\
\end{array}
\right)\left(
\begin{array}{c}
 \tilde{A}^{df}(1234) \\
 \tilde{A}^{df}(1243) \\
 \tilde{A}^{df}(1324) \\
\end{array}
\right).
\end{align}
\end{widetext}
This demonstrates that the KLT-like double-copy structure is a direct consequence of the underlying algebraic duality for any valid color structure.

\subsection{The symmetric ($dd$) sector}

The symmetric sector is built from color factors $c^{dd}(1234) = d^{12e}d^{e34}$. These factors are symmetric under leg exchanges within each vertex, e.g., $c^{dd}_s(1234)=c^{dd}_s(2134)$, and satisfy no Jacobi-like relations among themselves. The three graph structures $c^{dd}_s, c^{dd}_t, c^{dd}_u$ form a linearly independent basis.
Here are the minimum set of numerator relationships for a dressing $n^{(dd)}$  to be considered color-dual to $c^{dd}$:
\begin{align}  
  \label{eq:ddColorDual_iso}  n^{dd}(i j;k l) &= n^{dd}(k l;i j)\, \\
\label{eq:ddColorDual_sym}  n^{dd}(i j;k l) &= n^{dd}(j i;k l)
\,.
\end{align}
The same comments regarding functional symmetry apply --- this is only relevant functionally for identical particle types under little group.  Functional symmetry across different channels follows again the similar discussion, where additional particle types may require the introduction of distinct graphs amongst the channels.  For the all-vector cases discussed here, functional isomorphism holds, and one can impose:
\begin{align}
 n^{dd}_s &= n^{dd}(12;34) \,,\\
 n^{dd}_t &=  \left. n^{dd}_s \right |_{2\leftrightarrow 4}= n^{dd}(14;32) \,, \\
 n^{dd}_u &= \left. n^{dd}_s \right |_{1\leftrightarrow 4}=  n^{dd}(42;31) \, .
\end{align}

 Amplitudes in this sector are of the form,
\begin{align}\label{eq:dd-amp}
    \mathcal{A}^{dd}
    &= \sum_{i \in \{s,t,u\}} \frac{c^{dd}_i n^{dd}_i}{P_i}\,,
\end{align}
where color-kinematics duality requires the numerators $n^{dd}_i$ to share the same manifest vertex symmetry as the color factors. The symmetric double copy~\cite{Carrasco:2022jxn} is then
\begin{align}\label{eq:dd-double-copy}
    \mathcal{M}^{dd}
    &= \sum_{i \in \{s,t,u\}} \frac{\tilde{n}^{dd}_i n^{dd}_i}{P_i}\,.
\end{align}

For certain gauge groups one can go to the trace basis. Unlike the case for the $ff$ and $fd$ dressings, an amplitude dressed with $dd$ in $SU(N_c)$ leads to both single-trace and double-trace terms when expressed in the trace basis. More explicitly, $\mathcal{A} ^{dd}$ can be written as
\begin{align}\label{eq:dd-amp-2}
    &\mathcal{A}^{dd }(1^{a_1} 2^{a_2} 3^{a_3} 4^{a_4})
   =\sum_{\sigma \in S_4 /Z_4} \text{Tr} \left( T^{\sigma(1)} \dots T^{\sigma(4)} \right) A^{dd} _{(0)}\left( \sigma \right)\nn
   &+\frac{1}{N_c}\sum_{\sigma \in S_4 /Z_4} \text{Tr} \left( T^{\sigma(1)} T^{\sigma(2)} \right) \text{Tr} \left( T^{\sigma(3)}  T^{\sigma(4)} \right)  A^{dd} _{(1)}\left( \sigma\right).
\end{align}
We see that the double trace terms are suppressed by a factor of $\frac{1}{N_c}$.  These terms vanish when we are dealing with the $N_c \rightarrow \infty$ limit. The coefficients of the trace basis are given by
\begin{align}\label{eq:dd-color-ordered-amp}
    A^{d d} _{(0)}(1234)&=\frac{n^{dd}(1234)}{s}+\frac{n^{dd}(2341)}{t},\nn
    \qquad A^{d d} _{(1)}(1234)&=-4\frac{n^{dd}(1234)}{s}.
\end{align}
It is easy to see that the coefficients of the single trace elements, $A^{dd} _{(0)}$, can be expressed in terms of the coefficients of the double trace elements,  $A^{dd} _{(1)}$, and vice versa.

In terms of the color-ordered amplitudes, we can write the double copied amplitude $ \mathcal{M}^{dd}(12 3 4)$ as 
\begin{widetext}
\begin{align}\label{dd-KLT}
    \mathcal{M}^{dd }(1 2 3 4) &=\frac{1}{2} \left(
\begin{array}{c c c}
 A^{dd} _{(0)}(1234) &
 A^{dd} _{(0)} (1243) &
 A^{dd} _{(0)} (1324)\\
\end{array}
\right) \left(
\begin{array}{ccc}
 0 & s & t \\
 s & 0 & u\\
 t & u & 0  \\
\end{array}
\right)\left(
\begin{array}{c}
 \tilde{A}^{dd} _{(0)}(1234) \\
 \tilde{A}^{dd} _{(0)} (1243) \\
 \tilde{A}^{dd} _{(0)} (1324)\\
\end{array}
\right).
\end{align}
\end{widetext}
We note that the absence of three-point vector amplitudes that are color dual to
$d^{abc}$ is not in conflict with the existence of factorizable $f d$ or $d
d$ structures: there exist factorization channels corresponding to
products of scalar-gluon amplitudes.

\subsection{The permutation invariant ($d_4$) sector}

Finally, the four point contact amplitudes for bosons are naturally represented in terms of structures that are color-dual to the permutation invariant $d^{a_1 a_2 a_3 a_4}=\tfrac{1}{6}\sum_{\sigma\in S_3}  \Tr(T^{a_1} T^{\sigma_2}T^{\sigma_3}T^{\sigma_4})$.
Here are the requirements for $n^{d_4}$ to be color-dual to $d^{a_1 a_2 a_3 a_4}$,
\begin{equation}
  n^{d_4}(1234) = n^{d_4}(\rho_1 \rho_2 \rho_3 \rho_4) \qquad \forall \qquad  \rho \in S_4\,.
  \label{eq:d4ColorDual}
\end{equation}
These statements again hold functionally only for all permutations relating same-little-group-type particles.  For all vector particles and the representations discussed in this section and in the ancillary files, this holds functionally for all permutations.  For mixed scalar gluon amplitudes the permutation invariance holds only within little-group subsectors.  Note that any function that satisfies \cref{eq:d4ColorDual} automatically satisfies the $(dd)$ color-dual relations of \cref{eq:ddColorDual_iso,eq:ddColorDual_sym}.  Strictly speaking this means a generative basis of $(dd)$ color-dual numerators must span all $d^{a_1 a_2 a_3 a_4}$ color-dual numerators, a point we comment on below.

The four-point amplitude when the graphs are dressed with $d_4$ can then be written as:
\begin{align}\label{eq:d4-amp}
    \mathcal{A} ^{d_4}(1^{a_1} 2^{a_2} 3^{a_3} 4^{a_4})
    &= d^{a_1 a_2 a_3 a_4} n^{d_4}(1234).
\end{align}

The color-weight $d^{a_1 a_2 a_3 a_4}$ is completely permutation invariant. A kinematic numerator  $n^{d_4}$, in order to be color-dual to $d^{a_1 a_2 a_3 a_4}$, must satisfy all permutations compatible with the little-group structure of its labels.

We can compose to $n^{d_4}$ from each of the above representations.    
\begin{align}\label{eq:d4-comp-rules}
a^{d_4} &= n^{ff}_s \tilde{n}^{ff}_s + n^{ff}_t \tilde{n}^{ff}_t + n^{ff}_u \tilde{n}^{ff}_u  &(ff \to d_4 \text{ composition})\\
a^{d_4} &= n^{dd}_s \tilde{n}^{dd}_s + n^{dd}_t \tilde{n}^{dd}_t + n^{dd}_u \tilde{n}^{dd}_u  &(dd \to d_4 \text{ composition})\\
a^{d_4} &= n^{df}_s \tilde{n}^{df}_s + n^{df}_t \tilde{n}^{df}_t + n^{df}_u \tilde{n}^{df}_u  + (d\leftrightarrow f)  &(df \to d_4 \text{ composition})
\end{align}
For local\footnote{No ``denominators'' in the numerators.} dressings, as we consider here, these  relations manifestly promote  $(ff)$, $(dd)$, and $(df)$ kinematic dressings to local $n^{d_4}$ dressings.

Consider the primary ingredient per copy of the symmetric KLT kernel for $ff$,
\begin{align} s t A^{ff}(1234) &= n^{ff}_s t + n^{ff}_t s  \\
&= n^{ff}_s (t-u) + n^{ff}_t (s-u) + n^{ff}_u (t-s) \\
&\equiv n^{d_4}(1234)
\end{align}
Note that $(t-u)$ is the color-dual $s$-channel dressing for the minimally coupled  (covariantized-free) scalar amplitude with all external scalars, and the others follow by relabeling. 

\subsection{Iterated LEGO $N$-copy}

As discussed in Ref.~\cite{Carrasco:2025pmi,Brown:2025xlo} a consequence of constructing spin-statistics satisfying gauge-invariant building blocks, e.g. four-point LEGO blocks like $s t A^{ff}(1234) = n^{ff}_s t + n^{ff}_t s$ (where the gauge theory is antisymmetric-adjoint color-dual) is the possibility to combine them to build amplitudes associated with arbitrary spin particles ---  appropriately symmetrized to satisfy all spin-statistics and associated higher-spin Ward identities as per Rarita-Schwinger.  This is  phenomenologically relevant in the context of massive higher-spin states associated with bound states.  In this current work we demonstrate the utility of considering the triple-copy in the massless gravitational case to span all four-point gravitational higher-derivative amplitudes related to the $R^3$ Lovelock term.

\subsubsection{A note about nomenclature:} 

In the LEGO framework of Ref.~\cite{Carrasco:2025pmi}, it is natural to
distinguish between building blocks that modify little-group weight, called
spinor-blocks, and building blocks that only modify mass dimension, called
scalar-blocks.  When we refer to  $N$-copy in the LEGO sense we are discussing
the number of distinct gauge-invariant blocks of nontrivial little-group
weight.  

This means, for example, in the LEGO sense we may refer to vector
theories like Yang-Mills and Born-Infeld as single-copies, tensor theories like
Einstein-Hilbert gravity as a double-copy, and scalar theories like the
biadjoint scalar theory, the nonlinear-sigma Model, and the special Galleon as
zero-copies.   This is distinct from the typical antisymmetric adjoint
description that might either recognize all of them as double-copies, or refer
to only the theories that have nontrivial kinematic weights (independent of
little-group weight) in both copies as double-copies which would classify: the
biadjoint scalar as a zero-copy, NLSM and YM as single copies, and special
Galleon, Born-Infeld, and Einstein-Hilbert gravity as double-copies.  

Indeed, especially in the context of $Z$-theory, the term triple copy has already been used in the literature (cf.~for example ref.~\cite{Azevedo:2018dgo}) where one of the copies invariably consists entirely of scalars, and only two of the copies contain nontrivial little-group weight. As we discuss in \cref{sec:grav_reps}, this has a distinct structure from the LEGO triple-copy, in that it can be represented in terms of a product of two gauge theory weights where one weight has taken on arbitrarily high mass-dimension.

\subsubsection{What necessitates the  $N$-copy for max spin-2}
The $N$-copy is only used where each copy has nontrivial little-group weight
(spin) for at least one particle.  While we compose to build blocks of higher spin, such composed blocks are not considered primary.  As we will show at four-points it is possible to have a gravitational four-point amplitude that requires three copies of (Vector $\oplus$ Scalar) intertwined.  There are valid linearly diffeomorphic amplitudes at four-points that contain terms like,
\begin{equation}
\label{eq:dangerous}
\text{Not Double Copy} =  (\pol_1 \cdot \pol_2) (\pol_2 \cdot \pol_3) (\pol_1 \cdot \pol_3)(\pol_4 \cdot k_1)^2\,.
\end{equation}
Any attempt to break this into a double-copy has one side already containing a graviton (at least one $\pol_i^\mu \pol_i^\nu$).  To break into gauge theory requires a product of three terms each separately belonging to linearly gauge-invariant amplitudes.  One such partitioning could be, 
\begin{equation} 
\underbrace{(\pol_1 \cdot \pol_2)}_{\text{VVSS}} \otimes    \underbrace{ (\pol_2 \cdot \pol_3)(\pol_4 \cdot k_1)}_{\text{SVVV}} 
\otimes    \underbrace{(\pol_1 \cdot \pol_3) (\pol_4 \cdot k_1)}_{\text{VSVV}} 
\end{equation}
 It turns out four-point gravitational amplitudes of the form \cref{eq:dangerous} at any mass dimension can all be described in terms of a basis element involving the $R^3$ Lovelock gravitons, which is nonvanishing only in $D > 6$, as we will discuss --- and it is precisely that basis element that requires a triple-copy at four points. 

Note, importantly that  individual terms like \cref{eq:dangerous} can be ``gauged away'' in particular dimensions on shell (and of course is not gauge invariant by itself). It is the presence of braided terms like the permutations over \cref{eq:dangerous} in generic spacetime dimensions, ultimately due to the Lovelock term, that pushes us to consider the triple copy for gravitons.

\subsection{Redundancies and the Minimal Basis}

While the universal double copy is defined functionally over all sectors, not all color-dual sectors are
independent. As mentioned, any functional numerator dual to $d^{a_1 a_2 a_3 a_4}$ [satisfying \cref{eq:d4ColorDual}] satisfies the relations necessary to be considered $(dd)$-dual at the kinematic level [\cref{eq:ddColorDual_iso,eq:ddColorDual_sym}], and therefore lies within the space of numerators spanned by a complete basis of ($dd$) color-dual kinematic weights (a point we verify explicitly in the guide to ancillary files).
 Therefore, a complete basis of independent four-point universal double-copy constructions is spanned by just three sectors: the traditional antisymmetric-adjoint ($ff$), the mixed-symmetric ($df$), and the purely symmetric ($dd$) sectors.  In addition, we find that the space of four-point gravitational amplitudes built out of $\mathcal{M}^{df}$ is already spanned by that obtained from $\mathcal{M}^{dd}$.


\section{Classifying color-dual numerators at four points}
\label{sec:classifying-gauge-theory}

We begin by constructing a complete basis of color-dual gauge-theory numerators. At four points, higher-derivative operators introduce an infinite tower of possible numerators, distinguished by their mass dimension which is given by the number of momenta appearing in their expressions. Our goal is to classify this infinite set in terms of a finite basis of equivalence classes.

\subsection{Method of classification}
Following the approach of Ref.~\cite{Carrasco:2019yyn}, we define an equivalence
relation that identifies numerators that are physically indistinct up to their
overall scaling with kinematic invariants. We begin by equating two numerators
when their corresponding partial amplitudes are physically equivalent.
For the traditional $ff$ sector, two numerators $n_1^{ff}$ and $n_2^{ff}$ are
equivalent if their color-ordered amplitudes are identical,
\begin{align}\label{eq:num-equiv-brute}
    n_1 ^{ff}\sim n_2 ^{ff} \quad \Leftrightarrow \quad A_1^{ff} = A_2^{ff},
\end{align}
since any differences between the numerators must be pure gauge terms.

To create a finite basis, we generalize this to an equivalence class under multiplication by any fully permutation-invariant polynomial $\mathcal{P}(s,t,u)$ of the Mandelstam variables. Two numerators belong to the same equivalence class if their partial amplitudes are related by such a polynomial:
\begin{align}\label{eq:num-equiv}
    n_1 \sim n_2 \quad \Leftrightarrow \quad A_1 = \mathcal{P}(s,t,u) A_2.
\end{align}
This relation allows us to identify a finite number of modular building blocks, from which all higher-dimension numerators can be generated by multiplication with permutation-invariant scalars. This reduces the problem of classifying an infinite tower of operators to the finite problem of finding these core building blocks.

\subsection{Basis of color-dual vector numerators}

We perform this classification for vector numerators in each of the color-dual sectors. A comprehensive search identifies a finite basis of numerator blocks for each algebraic structure. The number of new, independent building blocks appearing at each mass dimension is summarized in \cref{table:amp-table}.

\begin{table}[h!] 
  \caption{Number of new, independent, vector numerator building blocks (generating independent on-shell amplitudes under scalar permutation invariants)  appearing at each mass dimension for the different color-dual sectors. The total gives the size of the finite basis for each sector.}
  \label{table:amp-table}
  \begin{ruledtabular}
  \begin{tabular}{lcccc}
    \shortstack{Numerator mass\\dimension} & $ff$ & Mixed ($df$)& $dd$ & $d_4$ \\
    \hline
    2  & 1 & 0 & 0 & 0 \\
    4  & 3 & 0 & 4 & 2 \\
    6  & 3 & 0 & 7 & 3 \\
    8  & 1 & 1 & 7 & 2 \\
    10 & 0 & 1 & 3 & 0 \\
    12 & 0 & 1 & 0 & 0 \\
    14 & 0 & 0 & 0 & 0 \\
    \hline
    Total & 8 & 3 & 21 & 7 \\
  \end{tabular}
  \end{ruledtabular}
\end{table}

This classification was previously carried out for the $ff$ sector in Refs.~\cite{Bern:2017tuc,Carrasco:2019yyn}, yielding a basis of eight numerators. Our analysis extends this to the other color structures, providing the complete set of gauge-theory inputs for the universal double copy. The basis for numerators dual to the quartic $d^{a_1 a_2 a_3 a_4}$ color factor is also included for completeness, although as discussed previously, these structures are redundant.

\subsection{Properties of generalized numerators}

A noteworthy property distinguishes numerators in the generalized sectors from
their traditional adjoint counterparts. Numerators dual to $df$ structures, and
likewise those dual to $dd$, are gauge invariant on their own, graph by graph.
This is in contrast to $ff$ numerators, which can be gauge-dependent,
with only their specific sum in a color-ordered amplitude being gauge invariant.

The graph by graph gauge invariance of the ($dd$) numerator directly follows from the fact that the ($dd$) color factors for the graphs are all independent of each other. Each of $c^{dd}_s$, $c^{dd}_t$, and $c^{dd}_u$ point to a different direction in color space.  Once we have expressed all of an amplitude's color factors --- including any contact contributions --- in terms of $c^{dd}_s$, $c^{dd}_t$, and $c^{dd}_u$, and collect on the color factors, we can examine their kinematic coefficients. If the amplitude is to be gauge invariant then those kinematic coefficients, the $n^{dd}_i/P_i$, must individually be gauge invariant.  Since the $P_i$ are just propagators, the $n^{dd}_i$ must be gauge invariant by themselves.

The graph by graph gauge invariance of the numerators for the ($df$) sector might be seen as surprising --- after all they satisfy a Jacobi relation that would seem to allow gauge-cancellation to creep in as it can  for the ($ff$) sector.  We can understand the graph-by-graph gauge invariance of the ($df$) sector  by considering the relationship between the full amplitude and the basis of numerators:
\begin{widetext}
\begin{equation}    
\mathcal{A}^{ df }(1^a 2^b 3^c 4^d)
    = \left(
\begin{array}{c c c}
 n^{df}_{1234} &
 n^{df}_{1324} &
 n^{df}_{1423} \\
\end{array}
\right)\left(
\begin{array}{ccc}
 -\frac{1}{2} \frac{\sigma_2}{\sigma_3}& \frac{1}{t}& -\frac{1}{u} \\
\frac{1}{t} &  -\frac{1}{2} \frac{\sigma_2}{\sigma_3} &\frac{1}{s}\\
 -\frac{1}{u} & \frac{1}{s}&  -\frac{1}{2} \frac{\sigma_2}{\sigma_3} \\
\end{array}
\right)\left(
\begin{array}{c}
 c^{df}_{1234} \\
 c^{df}_{1324} \\
 c^{df}_{1423}\\
\end{array}
\right). 
\label{eq:df-KLT-inverse}
\end{equation}
\end{widetext}
We introduce permutation invariants $\sigma_n \equiv s^n+t^n+u^n$.
Since the full amplitude $\mathcal{A}^{df}$ must be
gauge invariant, and the color structures in the final column vector are
linearly independent, each coefficient of a color factor --- which is a linear
combination of the $n^{df}_i$ --- must be gauge invariant. Because the inverse momentum-kernel matrix
$\mathcal{K}^{-1}$ is invertible, this implies that each numerator
$n^{df}_i$ must be individually gauge invariant. The analogous argument  applies directly to
the $dd$ sector. 

This analogous argument does not apply to $ff$ numerators because their
corresponding inverse kernel is singular, reflecting the linear dependence
(Jacobi identity) of the color factors.

\section{Higher-Derivative $D$-dimensional Gravity Amplitudes}
\label{sec:classifying-gravity}
Having classified the gauge-theory inputs, we now turn to the other side of the double-copy duality: the space of four-graviton amplitudes that the universal double copy aims to construct. Our goal is to perform a complete, first-principles classification of all possible four-graviton amplitudes, against which we can compare the output of our double-copy construction.   It is important to note that we classify the basis of operators for general spacetime dimension $D$, retaining terms that would otherwise vanish in $D=4$.

\subsection{Constraints from physical principles}

At the field-theoretic level, gravitation is understood as a theory of massless spin-$2$ particles  invariant under spacetime diffeomorphism. We can interpret the spin-2 field, $h_{\mu \nu}$, as parametrizing the fluctuations of the spacetime metric about the flat space:
\begin{align}
    g_{\mu \nu} = \eta_{\mu \nu}+\kappa h_{\mu \nu},
\end{align}
where $g_{\mu \nu}$ describes the full space-time metric and $\kappa$ is the coupling constant that can be expressed in terms of Newton's gravitational constant $G_N$ as $\kappa = \sqrt{32 \pi G_N}$.  Note that we are not necessarily taking $h$ to be a small fluctuation, simply the deviation from flat space-time.  In the context of scattering we are assuming an asymptotically flat background.

The linearized diffeomorphism invariance of the theory allows us to describe the dynamics of the theory in terms of that of a symmetric traceless tensor, which is consistent with the field-theoretic interpretation of an irreducible representation of a massless spin-2 particle as a symmetric traceless tensor.

 We can take the external graviton states to be $\pol_i ^{\mu \nu} \equiv \pol_i ^{((\mu} \pol_i ^{\nu))}$, where $\pol^{\mu}$ is a gauge-theory polarization vector, and the double parenthesis indicates an explicit symmetrization and removal of trace.  Such an external state transforms in the symmetric traceless representation of the little group, by construction. The gluon polarization vector transforms under a gauge transformation as 
 \begin{align}
     \pol^{\mu} \rightarrow \pol^{\mu}+p^{\mu}.
 \end{align}
 As such, we require our external graviton state to transform as
\begin{align}\label{eq:graviton-diffeomorphism}
  \pol^{\mu \nu} = \pol^{\mu } \pol^{\nu }&\rightarrow (\pol^{\mu}+k^{\mu})(\pol^{\nu}+k^{\nu})
  \nn
  &=\pol^{\mu \nu}+k^{\mu}\pol^{\nu}+k^{\nu}\pol^{\mu}+k^{\mu}k^{\nu}.
\end{align}
 This is equivalent  to the statement of linearized diffeomorphism invariance for the graviton state, which in general is given by
 \begin{align}
     \pol^{\mu \nu} \rightarrow \pol^{\mu \nu} + k^{\mu} q^{\nu}+ k^{\nu} q^{\mu},
 \end{align}
 where $q$ is an arbitrary reference vector that satisfies $ k \cdot q = 0$. Indeed we recover \cref{eq:graviton-diffeomorphism} by setting $q^{\mu}=\pol^{\mu}+\frac{1}{2}k^{\mu}$.

Bose symmetry dictates that the gravitational amplitude $\mathcal{M}$ must be invariant under the permutation of external particle labels. As a tree-level quantity, $\mathcal{M}$ is a rational function of Lorentz invariants, where the poles correspond to physical factorization channels. To systematically classify the complete space of such functions  we introduce the \emph{universal numerator, $\mathcal{N}$}, defined such that the full amplitude is given by
\begin{align}
    \mathcal{M} = \frac{\mathcal{N}}{stu}\,.
\end{align}
Here, the denominator $stu=\frac{1}{3}\sigma_3$ is the unique permutation-invariant scalar of mass
dimension 6. We define $\mathcal{N}$ as the space of local polynomials in
Lorentz invariants of momenta and polarization vectors that satisfies three conditions: it must be fully permutation invariant, linear in graviton polarization tensors, and it must be invariant under the linearized diffeomorphism symmetry (gauge invariance) of the external gravitons.

This definition allows us to span the entire space of possible gravity amplitudes---both factorizable and contact terms---by classifying the vector space of polynomials $\mathcal{N}$ at each mass dimension. This unifies the treatment of standard (and exotic) massless exchange as well as higher-derivative operators.  A pure contact term, being a polynomial with no physical poles, corresponds to a universal numerator $\mathcal{N}$ that contains an overall factor of $stu$, ensuring the cancellation of the denominator. While the number of physically distinct factorization channels for massless spin-2 exchange is finite\footnote{Assuming we cap the highest spin exchange of massless particles at spin-two as per Ref.~\cite{Weinberg:1965nx}.}, the tower of possible higher-derivative contact terms at four-points is infinite. By focusing on $\mathcal{N}$, we capture all such contributions without needing to separate the basis into distinct sectors for exchange diagrams versus local contact operators. Any valid physical amplitude simply corresponds to a specific element within $\mathcal{N}$.

The power of this framework is that it reveals an important structure within the infinite tower of possible gravitational operators. The full space $\mathcal{N}$ is infinite-dimensional, but it is not arbitrary. It is generated by a finite set of unique building blocks.  In our analysis we identify a basis of \emph{fundamental gravitational polynomials}, denoted $\mathcal{G}$, which are the unique gravitational structures that cannot be generated by multiplying a lower-dimension polynomial by a scalar permutation invariant. By construction, we project out the contributions of $\mathcal{P}(s,t,u) \mathcal{G}^{(m)}$ from $\mathcal{G}^{(n>m)}$. The complete space $\mathcal{N}$ is then the space spanned by these gravitational fundamentals, $\mathcal{G}$, and their products with all scalar permutation invariants $\mathcal{P}(s,t,u)$. In essence, any possible higher-derivative interaction at any mass-dimension can be expressed as a linear combination of these fundamentals weighted by scalar permutation invariants.

We summarize the classification of both the full space and the fundamental basis in \cref{table:grav-amp-table}. The second column shows the number of linearly independent universal numerators at mass dimension $d$, denoted $|\mathcal{N}^{(d)}|$. The third column lists the number of new gravitational fundamentals that appear at that dimension, $|\mathcal{G}^{(d)}|$. The first nonvanishing diffeomorphism-invariant polynomials appear at mass dimension 6. The number of universal numerators we obtain is consistent with those obtained from the partition function for polynomial $S$-matrices given in Ref. \cite{Chowdhury:2019kaq}. Strikingly, our analysis reveals that the basis of these fundamental polynomials is finite; no new structures appear beyond mass dimension 14.   

Where do familiar gravity amplitudes lie in this classification? The
Einstein-Hilbert amplitude is of mass-dimension two, which corresponds to a
universal numerator $\mathcal{N}^{(8)} = stu \mathcal{M}^{\text{GR}}$ at mass
dimension eight. The leading higher-derivative correction, a single insertion of
$R^3$, corresponds to a factorizing amplitude at four-point. In our basis, this
appears as a universal numerator of mass dimension 12 (represented by $\mathcal{N}^{(12)} = stu \times \mathcal{M}^{R^3}$).

\begin{table}[t]
  \centering
  \caption{At each mass-dimension $d$, the total number of parity-even universal gravitational numerators  $\mathcal{N}^{(d)}$ in generic spacetime dimension $D$, and the number of new  gravitational fundamental numerators  $\mathcal{G}^{(d)}$ (independent under multiplication by scalar permutation invariants). }
  \label{table:grav-amp-table}
  \begin{ruledtabular}
    \begin{tabular}{ccc}
    \shortstack{Mass dimension\\$(d)$} 
        & \shortstack{Universal numerators\\ $|\mathcal{N}^{(d)}|$}
        & \shortstack{Gravitational fundamentals\\ $|\mathcal{G}^{(d)}|$}
        \\
      \hline
       0  & 0  & 0    \\
       2  & 0  & 0    \\
       4  & 0  & 0    \\
       6  & 1  & 1  \\
       8  & 6  &  6  \\
      10  & 10  & 9  \\
      12  & 17 &  10  \\
      14  & 19 & 3 \\
      16  & 26 & 0  \\
      18  & 30 & 0   \\
    \end{tabular}
  \end{ruledtabular}
\end{table}


\section{Spanning Gravity with the Universal $N\leq3$ Copy}
\label{sec:factorize-gravity}

Having independently classified the basis of color-dual gauge-theory numerators (Sec.~\ref{sec:classifying-gauge-theory}) and the basis of fundamental gravitational polynomials $\mathcal{G}$ (Sec.~\ref{sec:classifying-gravity}), we are now in a position to relate them. We will now demonstrate that the full space of four-graviton amplitudes is constructable by gauge theory building blocks.

The double copy provides a concrete reductive spin-addition prescription for fundamental blocks. For four-point amplitudes involving external spin-2, these constructions fall into two primary classes, distinguished by their lowest weight bosonic building blocks:
\begin{enumerate}
    \item \textbf{Four-gluon double copy:} Amplitudes built from two four-gluon theories, schematically represented as $(gggg) \otimes (gggg)$. These can be constructed using $ff$ and $dd$ double-copy of \ \cref{eq:ff-KLT,dd-KLT}.  The kinematic space spanned by the $df$ double-copy here is a subspace of that generated by $dd$ and $ff$.
    \item \textbf{Scalar-gluon triple copy:} Amplitudes built from theories with both scalars and gluons, such as $ (ggss) \otimes (sggg) \otimes (gsgg) +\text{perms}$.  This can be interpreted in the LEGO sense of $N$-copy to construct higher-spin amplitudes.
\end{enumerate}
We find that these two construction types are sufficient to span the entire
basis $\mathcal{G}$. We note that the enumeration listed above exhausts all the
ways a tensor with the external states of four spin-2 particles can be composed
of products of tensors corresponding to bosons with spin less than 2. When the
spacetime dimension $D \leq 6$, the four-gluon double copy is sufficient to fully span all graviton fundamentals $\mathcal{G}$. The scalar-gluon triple copy is only relevant for graviton amplitudes in $D >6$.

One might be curious about the relevance higher $N$-copies at four points.  One can construct them, but they are all spanned by the two above classes.  For example  $(ggss)\otimes(gssg)\otimes(ssgg)\otimes(sggs)$ is spanned by the double copy,
\begin{equation}
 (gggg) \sim (ggss) \otimes (ssgg) \sim (gssg)\otimes (gssg).
\end{equation}

While we do not list the explicit functional forms of the higher-dimension numerators here due to their length, we provide the complete machine-readable expressions for the basis of gauge-theory numerators and the resulting gravitational fundamentals in the ancillary files attached to this submission. We demonstrate explicitly that every four-point  gravitational fundamental is spanned in the guide to ancillary files.

\subsection{The four-gluon contribution}

We begin by systematically constructing all possible four-graviton amplitudes arising from the double copy of our complete basis of color-dual four-gluon numerators. This corresponds to the $(gggg) \otimes (gggg)$ class. Specifically, we compute the set of all amplitudes $\{\mathcal{M}^{ff}, \mathcal{M}^{dd}\}$ by taking the numerators from \cref{table:amp-table} as inputs. We exclude the $\mathcal{M}^{df}$ construction, as its gravitational output is linearly dependent on the amplitudes generated by the $\mathcal{M}^{dd}$ sector.

Upon comparing this generated set of amplitudes to our basis of fundamental gravitational polynomials, we find a remarkable correspondence. All fundamental polynomials $\mathcal{G}_i$ with a mass dimension of eight or higher are spanned by these double copies of parity-even four-gluon interactions\footnote{We remind the reader that the projection of $\mathcal{G} ^{(n >6)}$ along $\mathcal{G} ^{(6)}$ is vanishing, by construction.} . This demonstrates that the vast majority of the gravitational EFT landscape is dictated by the $ff$ and $dd$ kinematic algebras. In $D \leq 6$, the graviton fundamental $\mathcal{G}^{(6)}$ is vanishing, and the $ff$ and $dd$ double copies span the entire space of four-graviton amplitudes at arbitrary mass dimension.
 
\subsection{Unpacking the triple copy}

Thus, for $D > 6$, the parity-even four-gluon double copy is not the complete story. The single fundamental gravitational polynomial at mass dimension 6, $\mathcal{G}^{(6)}$, and the infinite tower of universal numerators it generates via multiplication with scalar permutation invariants, are absent from the parity-even $(gggg) \otimes (gggg)$ construction in arbitrary dimensions. The failure of double copies of $(gggg) \otimes (gggg)$ to span $\mathcal{G}^{(6)}$ can be traced to the presence of terms of the form \cref{eq:dangerous}
in 
\begin{equation}
\mathcal{G}^{(6)}
\propto
\,\delta^{\mu_1}_{[\nu_1}\cdots\delta^{\mu_7}_{\nu_7]}
\pol_1{}_{\mu_1}\pol_2{}_{\mu_2}\pol_3{}_{\mu_3}\pol_4{}_{\mu_4} k_1{}_{\mu_5} k_2{}_{\mu_6} k_3{}_{\mu_7} 
\pol_1^{\nu_1}\pol_2^{\nu_2}\pol_3^{\nu_3}\pol_4^{\nu_4}
k_1^{\nu_5}k_2^{\nu_6}k_3^{\nu_7}\, .
\label{eq:delta_G6}
\end{equation}
The antisymmetrized products of $\delta^{\mu \nu}$s can be written in terms of the seven-dimensional
Levi-Civita symbol: $\levi^{\mu_1 \dots \mu_7
} \levi_{\nu_1 \dots \nu_7}$, which makes it clear that $\mathcal{G}^{(6)}$ vanishes for spacetime dimension $D \leq 6$. 

This four point amplitude of \cref{eq:delta_G6} is generated by the  $R^3$ Lovelock term at cubic order \cite{Chowdhury:2019kaq}.   We should be very clear this is not the same $R^3$ operator which is the doublecopy of $F^3$ and contributes a non-vanishing three-point amplitude.  Lovelock terms are quite special and distinct, and worth a brief discussion.

Lovelock theories of gravity were originally studied in the context of classifying operators that lead to tensors that have at most two derivatives on each metric, are symmetric and are covariantly conserved. Such tensors lend themselves as natural candidates for generalizations of the Einstein tensor \cite{Lovelock:1971yv,Farhoudi:1995rc}. The properties that we listed above are restrictive enough that there exists a unique operator at each order in Riemann tensors, which is given by
\begin{align}
\sqrt{-g} \; \delta^{\mu_1} _{[\nu_1} \dots \delta^{\mu_{2n}} _{\nu_{2n}]} R_{\mu_1 \mu_2} ^{\; \; \ \; \; \; \nu_1 \nu_2} \dots R_{\mu_{2n-1} \mu_{2n}} ^{\; \; \; \; \; \; \nu_{2n-1} \nu_{2n}}. 
\end{align}
The Lovelock term at quadratic order gives us the celebrated Gauss-Bonnet term, whereas the Lovelock term at cubic order generates the fundamental polynomial $\mathcal{G}^{(6)}$ as a contact amplitude.

The operator corresponding to the Lovelock term at order $n$ is topological in
$D=2n$ and is vanishing for $D<2n$
\cite{Farhoudi:1995rc,Zwiebach:1985uq,Zumino:1985dp}. As such, the Gauss-Bonnet
term is purely topological in $D=4$, whereas the operator that produces
$\mathcal{G}^{(6)}$ is topological in $D=6$ and vanishing for $D < 6$. Moreover,
the Lovelock term at order $n$ has nonvanishing $S$-matrix elements only for
amplitudes relating to $(n+1)$ or more gravitons \cite{Zumino:1985dp}. This is
consistent with the fact that the Gauss-Bonnet term does not lead to any
modifications of the propagator. In addition, this would imply that the first
contribution of the cubic Lovelock term corresponds not to a cubic
interaction, but a quartic contact amplitude, consistent with our
interpretation of $\mathcal{G}^{(6)}$ as a contact amplitude.

It should be noted that in particular spacetime dimensions it is possible to ``gauge away'' all terms of the form \cref{eq:dangerous} on shell (i.e. eliminated by a judicious choice of polarization vectors).  For example one can write \cref{eq:delta_G6} as,
\begin{equation}
\mathcal{G}^{(6)}(1,2,3,4)
\propto \left( \pol_1 \wedge \pol_2 \wedge \pol_3 \wedge \pol_4 \wedge k_1 \wedge k_2 \wedge k_3 \right)^2
\end{equation}
In generic dimensions the square is performing seven Lorentz contractions, but in precisely $D=7$  there exists an antisymmetric  Levi-Civita symbol that can contract with a single copy of the wedge products to form a Lorentz scalar.  So precisely in seven dimensions each copy can be associated with a parity-odd vector amplitude, and we can interpret the parity-even $\mathcal{G}^{(6)}$ as a literal double copy.  While it is notable that there exists a parity-odd double copy in that particular dimension, when confronted with reproducing graviton contributions like \cref{eq:delta_G6} in arbitrary dimensions we now turn to the triple copy. 

Explicitly, the gravitational fundamental $\mathcal{G}^{(6)}$
can be decomposed as:
\begin{align}\label{eq:triple-copy}
    \mathcal{P}(s,t,u) \mathcal{G}^{(6)} = \mathcal{M}^{ff/dd}+\left( A^{d_4}(1^g 2^g 3^s 4^s)  A^{d_4}(1^s 2^g 3^g 4^g)  A^{d_4}(1^g 2^s 3^g 4^g) +\text{perms}\right) ,
\end{align}
where the partial amplitudes $A^{d_4}$ are those of a gauge theory with an adjoint scalar, obtained from one of the composition rules listed in \cref{eq:d4-comp-rules},  $\mathcal{M}^{ff/dd}$ is the set of gravity amplitudes obtained by performing an $ff$ or $dd$ double copy on two pure gluon amplitudes, and $\mathcal{P}(s,t,u)$ is a permutation-invariant function built out of Mandelstam invariants. The sum is over all $4!$ permutations of external leg labels. For concreteness, we can take $\mathcal{P}(s,t,u) = (stu) $. After projecting out the component of $(stu) \mathcal{G}^{(6)} $ that can be spanned by $ \mathcal{M}^{ff/dd}$, we are left with
\begin{align}\label{eq:min-trip}
 (stu)  \; \mathcal{G}^{(6)} \equiv \sum_{\text{perms}} \left( t u F_{12}    F_{234}  F_{134} \right) \qquad \text{mod  } \{ \mathcal{M}^{ff/dd}\},
\end{align}
where $F_i ^{\mu \nu} \equiv k_i ^{\mu} \pol_i ^{\nu}-k_i ^{\nu} \pol_i ^{\mu}$
is the linearized field-strength tensor and $F_{i_1 \dots i_n}\equiv
\tr\left(F_{i_1} \dots F_{i_n}\right)$. We can recognize the expression on the right-hand side of \cref{eq:min-trip} as arising from a triple copy as follows,
\begin{align}
&\underbrace{\left((s-t)^2(s-u)^2(t-u)^2 (s t u) \right) }_{\mathcal{P}(s,t,u)}   \mathcal{G}^{(6)}  \,\,\, \text{ mod  } \{ \mathcal{M}^{ff/dd}\}\nn
&=\underbrace{tuF_{12} }_{A^{d_4}(1^g 2^g 3^s 4^s)}  \underbrace{(s-t)(s-u)(t-u)F_{234}}_{A^{d_4}(1^s 2^g 3^g 4^g)} 
 \underbrace{(s-t)(s-u)(t-u)F_{134}}_{A^{d_4}(1^g 2^s 3^g 4^g)} +\text{perms} .
\end{align}
The sum over all permutations is required for the Bose-symmetry of the resulting gravitons.  

It should be noted that there are many ways of representing this contribution as a triple copy of nontrivial little-group weights, and it need not be in the form of kinematics dual to $(d_4)$.  In the guide to ancillary files notebook we present an additional representation in terms of ($ff$) building blocks.

Nevertheless, with this construction in hand, for general spacetime dimension $D$, the complete basis of fundamental gravitational polynomials $\mathcal{G}$ is spanned by the union of the standard $(gggg) \otimes (gggg)$ double copy and the $(ggss) \otimes (sggg) \otimes (gsgg) + \text{perms}$ triple copy.  This validates the generalized color-kinematics duality principle, showing that the entire landscape of four-point gravitational interactions can be factorized into gauge-theory building blocks, once admitting all allowed algebraic sectors.


\section{On the Representation of Gravitational Interactions}
\label{sec:grav_reps}
We have shown that the complete set of four-graviton amplitudes can be generated from a finite set of gauge-theory building blocks. In particular, the full space of parity-even gravitational amplitudes is spanned by the union of the double-copy construction, built from parity-even four-gluon amplitudes, together with a single additional structure arising from a triple copy involving scalar-gluon sectors, with the latter sector being redundant in $D  \leq 6$. This establishes that the entire four-point gravitational $S$-matrix can be obtained from gauge-theory data, upon allowing for a generalized notion of the duality between color and kinematics.

It is worth emphasizing that this result goes beyond the familiar realization of gravity and its UV completion in string theory as double copies of gauge theory. Perturbative string theory  furnishes a nontrivial example of double-copy: its tree-level amplitudes admit a representation in which closed-string amplitudes arise from the Kawaii-Lewellen-Tye string double copy of Chan-Paton stripped open-string amplitudes, encoding an infinite tower of higher-derivative interactions through $\alpha'$ corrections.  The Chan-Paton stripped open-string amplitudes do not satisfy field theory BCJ relations --- rather stringy generalizations known as monodromy relations.   Representing string orderings with capital indices, we can understand the closed string amplitudes as follows:
\begin{equation}
    \mathcal{A}^{\text{CS}} = A^{\text{OS}}_I (\otimes_{\alpha'})^{I J} A^{\text{OS}}_J,
\end{equation}
where $\otimes_{\alpha'}$ represents the $\alpha'$ dependent matrix combining various permutations of the orderings of stripped open-string amplitudes.

Each open-string amplitude is itself a field-theory double-copy of a gauge
theory (sYM for the super-string, YM+(DF)$^2$ for the Bosonic string) with
Z-theory, an all-order in $\alpha'$ nonlocal scalar EFT~\cite{Broedel:2013tta,
Carrasco:2016ldy,Carrasco:2016ygv, Mafra:2016mcc, Azevedo:2018dgo}.   
Z-theory
is a bicolored scalar theory, whose amplitudes are given by color-dressing
disk-integrals, where one color is simply antisymmetric adjoint $f^{abc}$ per
vertex but the other color carries Chan-Paton factors mixed with kinematics so
that both factors satisfy Jacobi and antisymmetry:
\begin{align}
\mathcal{Z} &= \sum_{g} \frac{z_g c_g}{P_g} \\
  c_i+c_j+c_k =0 &\iff z_i + z_j +z_k=0\,.
\end{align}
Stripping only the field-theory index yields the Chan-Paton dressed, field theory ordered amplitudes $\mathbf{Z}_i$.
We can write
\begin{equation}
\mathbf{\mathcal{A}}^{\text{OS}} _I= \mathbf{Z}^{\text{OS}}_{Ij}\otimes^{jk}  A_k = \sum_g \frac{z_g n_g}{P_g} 
\end{equation}
where the dressings $z_g$ and $n_g$ satisfy Jacobi about each edge and antisymmetry around each vertex, and $\otimes^{i j}$ represents the field theory KLT or momentum kernel.  The closed string is written as a field theory double-copy then as:
\begin{align}
 \mathcal{A}^{\text{CS}} &= (A_i \otimes^{i j}  Z_{Ij} )\otimes_{\alpha'} ^{I J} (Z_{Jk} \otimes^{k l}  A_l) \nonumber \\
 &= A_i \otimes^{i j}  (Z_{Ij} \otimes_{\alpha'} ^{I J} Z_{Jk} \otimes^{k l}  A_l)\nonumber
 \\ &= A_i \otimes^{i j} A^{\text{sv}}_j(\alpha') \nonumber \\
  &= \sum_g \frac{n_g \, n^{\text{sv}}_g(\alpha')}{P_g}, \\
   n_i+n_j+n_k =0 &\iff n^{\text{sv}}_i + n^{\text{sv}}_j +n^{\text{sv}}_k=0\,,
 \end{align}
 with $\mathcal{O}^{\text{sv}}$ representing an all-order in higher-derivative gauge interactions with single-valued multiple-zeta-value Wilson coefficients. Note that if $A_i$ satisfies field theory color-dual relations then $A^{\text{sv}}_i$ does as well, and thus can be written in terms of antisymmetric adjoint ($ff$) color-dual numerators. 
 We can interpret $Z \otimes_{\alpha'} Z$ as a generalized field-theory KLT kernel~\cite{Carrasco:2021ptp, Chi:2021mio, Chen:2023dcx}, or as one mechanism to achieve the type of exponentiation associated with with color-dual kinematics, factorization, and higher-derivative operators~\cite{Carrasco:2021ptp, Carrasco:2022lbm, Carrasco:2022sck, Brown:2023srz, Li:2023wdm, Carrasco:2023wib}.  Note in either framework there are only two copies with nontrivial little-group weight in the closed string construction.  

The present analysis shows that even though it is the only known consistent UV completion for gravitation, perturbative string theory as currently framed cannot be the complete story for all gravitational operators. The complete four-point gravitational effective S-matrix requires structures that cannot be obtained from only a double copy of gauge-theory amplitudes. In particular, the Lovelock $R^{3}$ interaction, which generates the unique parity-even dimension-six gravitational amplitude, cannot arise from any double-copy construction in a generic dimension. Instead, it is realized through a triple-copy structure,
\begin{equation*}
(\text{YM}+\text{scalar}) \;\otimes\; (\text{YM}+\text{scalar}) \;\otimes\; (\text{YM}+\text{scalar})\,,
\end{equation*}
with the physical graviton emerging from the appropriate projection onto the symmetric, traceless spin-2 sector. In this sense, the familiar Einstein-Hilbert interaction corresponds to the \((\text{spin-1})\otimes(\text{spin-1})\) channel, while the Lovelock term probes a genuinely higher-rank structure that lies beyond the reach of any two-fold copy.  String theory may yet be the only UV completion of gravity, but that will mean that Lovelock terms are excluded from UV complete theories.

This perspective suggests that the appearance of a triple-copy structure is not accidental but reflects a deeper organizing principle underlying the full possibility of gravitational effective field theories. The Lovelock interaction should therefore be viewed as the first signal of a more general hierarchy, in which increasingly intricate algebraic structures are required to realize the maximally entwined higher-derivative interactions encoded in higher multiplicity Lovelock terms. Whether this hierarchy is useful, terminates, or continues indefinitely remains an open question, but its existence strongly indicates that color-kinematics duality provides a unifying framework extending well-beyond the traditional double-copy paradigm.


\section{Conclusion}

In this work we have shown that the BCJ duality between color and kinematics, when extended to the full algebra of four-point color structures, provides a complete constructive principle for gravitational effective field theory. In particular, we have demonstrated that the entire space of parity-even four-graviton amplitudes is spanned by a finite set of gauge-theory building blocks, requiring at most a triple-copy structure, with the $R^3$ Lovelock term representing the boundary of double-copy construction.

This result establishes that the gravitational four-point S-matrix admits a fully gauge-theoretic origin: every allowed interaction can be expressed in terms of color-dual kinematic data. While the conventional double copy accounts for a large class of such amplitudes, the full structure requires the inclusion of additional algebraic sectors whose role becomes essential in reproducing the complete space of higher-derivative interactions.

The double copy is known to generate well-defined gravitational dynamics with constraining rigidity in higher-derivative corrections ---
including Einstein gravity and its string-theoretic extensions.  It is important to emphasize that the appearance of a triple-copy structure at this one mutliplicity does not, by itself, imply the existence of a consistent fundamental triple-copy theory.  While it is clear that there exists a consistent factorizable gravitational theory that includes the $R^3$ Lovelock term, here we have only shown that its four-point amplitude is described by a triple copy.  Additional analysis, which we leave to future work, is required to demonstrate whether or not every $m$-point tree-level amplitude in this theory can always be described by a triple copy between the same three gauge theories.

From this perspective, the role of the triple copy is not to introduce new
dynamical content, but to expose the minimal algebraic structure needed to
reproduce the full space of gravitational amplitudes at four points. The Lovelock term in isolation is poorly behaved in the UV in any dimension it can contribute. However, current-known bounds arising from causality do not fully exclude the presence of the cubic Lovelock term in potential UV completions of gravity \cite{Camanho:2014apa,Caron-Huot:2022ugt,Caron-Huot:2022jli}. It would be an interesting to understand if an infinite tower of higher-derivative operators could be introduced to resum to reasonable behavior in the UV --- an option closed string theory provides for double-copy consistent higher-derivative operators. We emphasize that the
triple copy structures are irrelevant to  higher-derivative four-graviton amplitudes 
in $D \leq 6 $, and the operators they are relevant for in $D > 6$ are excluded from 
the known UV completions of gravity in string constructions.

Finally, the results presented here suggest a broader organizing principle for gravitational interactions. The appearance of a finite set of fundamental building blocks at four points hints at a deeper structure governing higher-point amplitudes. Determining whether similar organizing principles persist at higher multiplicity, and how they relate to the universal notion color-kinematics duality, represents a natural and promising direction for future investigation.

\acknowledgments
We are especially grateful to Zvi Bern, Henrik Johansson, Radu Roiban, and Oliver Schlotterer for related collaboration, helpful discussion, and useful feedback on an earlier draft of this manuscript.
This work was supported by a grant from the Simons Foundation International
[SFI-MPS-SSRFA-00012751, A.E,].
This work was also supported in part by the DOE under Contract No. DE-SC0015910, and by Northwestern University via the
Amplitudes and Insights Group, Department of Physics and Astronomy, and
Weinberg College of Arts and Sciences.

\bibliographystyle{apsrev4-1} 
\bibliography{refs}

\end{document}